\documentclass[12pt]{article}

\begin{document}

\title{
{\Large \bf Generalized quantum field theory:
perturbative computation and perspectives }
}         

\author{V. B. Bezerra \\
Universidade Federal da Paraiba, Depto. de F\'\i sica \\
Caixa Postal 5008, 58051-970 - Jo\~{a}o Pessoa, PB, Brazil \\
e-mail: valdir@fisica.ufpb.br \\
and \\
E. M. F. Curado and M. A. Rego-Monteiro \\
Centro Brasileiro de Pesquisas F\'\i sicas, \\ 
Rua Xavier Sigaud 150, 22290-180 - Rio de Janeiro, RJ, Brazil\\
e-mail: eme@cbpf.br and regomont@cbpf.br \\
}        
\date{}          
\maketitle
\begin{abstract}
 
\indent

We analyze some consequences of two possible interpretations
of the action of the ladder operators emerging from generalized
Heisenberg algebras in the framework of the second quantized
formalism. Within the first interpretation we 
construct a quantum field theory that creates at
any space-time point particles described by a $q$-deformed
Heisenberg algebra and we compute the propagator and a specific
first order scattering process. Concerning the second one, we draw 
attention to the possibility of constructing this theory where 
each state of a generalized Heisenberg algebra is interpreted as a 
particle with different mass.

\end{abstract}

\vspace{1cm}

\begin{tabbing}

\=xxxxxxxxxxxxxxxxxx\= \kill


{\bf journal-ref. :} Phys. Rev. {\bf D 65} (2002) 065020.

\end{tabbing}

\newpage

\section{Introduction}

A class of algebras which is a generalization of the Heisenberg 
algebra has recently been constructed $^{\cite{algebra1, jpa}}$.
These algebras are characterized by functionals depending
on one of their generators, that are called the characteristic
functions of the algebra. When this functional is
linear with slope $\theta$ this algebra turns into a 
$q$-oscillator algebra $^{\cite{qosc}}$, being $\theta$ 
related to the deformation parameter as $q^2 = \tan \theta$. 
It is worth noticing that when this functional is
a general polynomial we obtain the so-called 
multiparametric deformed Heisenberg algebra.

In what concerns their physical interpretation, these
generalized algebras describe the Heisenberg-type algebras
of one-dimensional quantum systems having arbitrary
successive energy levels given by $\epsilon_{n+1}=
f(\epsilon_{n})$ where $f(x)$ is the characteristic
function of the algebra $^{\cite{comhugo}}$. As examples we 
can mention
the non-relativistic $^{\cite{comhugo}}$ and 
relativistic $^{\cite{campos}}$ square-well
potentials in one-dimension and the harmonic oscillator
on a circle $^{\cite{circulo}}$. It is also interesting
to mention that the representations of the 
algebra for a general characteristic function were
constructed studying the stability of the fixed points
of this function and of their composed functions $^{\cite{jpa}}$. 

Heisenberg algebra is an essential tool in the second 
quantization formalism
because their generators create and annihilate particle
states. In the generalized
Heisenberg algebras the generators of the algebra are
also ladder operators constructing all different
energy levels of a one-dimensional quantum system from
a given energy level. As in the generalized case the 
energy difference of any two successive levels is not
equal, two possible interpretations of the action of the
ladder operators come out. In the first interpretation 
we have one similar
to that in the harmonic oscillator case, but
with the difference that the total energy of $n$ particles is
not equal to $n$ times the energy of each particle while
in the second interpretation we associate particles with 
different masses to different energy levels. 

It then seems
natural to investigate the possibility of extending
these interpretations of the ladder operators of
generalized Heisenberg
algebras in the framework of the second quantized formalism.
In this paper we start the analysis of the harmonic 
oscillator-like interpretation. We construct a QFT 
having fields that produce at any space-time point,
particles satisfying a $q$-deformed Heisenberg
algebra and we start the analysis of the perturbation
theory for this case. The interest in this QFT comes from the fact
that creation and annihilation operators of correlated fermion pairs, in 
simple many body systems, satisfy a deformed Heisenberg algebra that can 
be approximated by $q$-oscillators $^{\cite{bonatsos}}$.

In section 2, we summarize the
generalized Heisenberg algebras; in section 3 we
implement a physical realization of a one-parameter
deformed Heisenberg algebra. In section 4, the quantum
field model following the harmonic oscillator-like
interpretation is presented. The propagator and a specific
first order scattering process are computed. In section 5, 
we discuss the possibility of constructing a QFT
where each state of its Hilbert space, which is created
by the ladder operators of the associated generalized
Heisenberg algebra, is interpreted as a particle 
with different mass. We end this section with some 
conclusions.

\section{Generalized Heisenberg algebras}

Let us consider an algebra generated by $J_{0}$, $A$ and $A^{\dagger}$ 
and described by the relations $^{\cite{jpa}}$
\begin{eqnarray}
J_{0} \, A^{\dagger} &=& A^{\dagger} \, f(J_{0}) ,
\label{eq:alg1} \\
A \, J_{0} &=& f(J_{0}) \, A , 
\label{eq:alg2} \\
\left[ A, A^{\dagger} \right] &=& f(J_{0})-J_{0} ,
\label{eq:alg3}
\end{eqnarray}
where $^{\dagger}$ is the Hermitian conjugate and, 
by hypothesis, $J_{0}^{\dagger}=J_{0}$ and $f(J_{0})$
is a general analytic function of $J_{0}$. 
It is simple to show that the generators of the algebra 
trivially satisfy the Jacobi identity $^{\cite{circulo}}$.

Using the algebraic relations in eqs. (\ref{eq:alg1}-\ref{eq:alg3}) we
see that the operator
\begin{equation}
C = A^{\dagger} \, A - J_{0} = A \, A^{\dagger} - f(J_{0})  
\label{eq:casimir}
\end{equation}
satisfies
\begin{equation}
\left[ C,J_{0} \right] = \left[ C,A \right] = 
\left[ C,A^{\dagger} \right] = 0  ,
\label{eq:comute}
\end{equation}
being thus a Casimir operator of the algebra.

We analyze now the representation theory of the algebra when 
function $f(J_{0})$ is a general analytic function of  
$J_{0}$. 
We assume we have an $n$-dimensional irreducible representation
of the algebra given in eqs. (\ref{eq:alg1}-\ref{eq:alg3}). 
We also assume that there is a state $|0\rangle$ with the lowest 
eigenvalue of the Hermitian operator $J_{0}$
\begin{equation}
J_{0} \, |0\rangle = \alpha_{0} \, |0\rangle .
\label{eq:alfa0}
\end{equation}
For each value of $\alpha_{0}$
we have a different vacuum, thus a better notation could
be $|0\rangle_{\alpha_0}$
but, for simplicity, we shall omit subscript $\alpha_0$.  

Let $| m \rangle$ be a normalized eigenstate of $J_{0}$, 
\begin{equation}
    J_{0} |m \rangle = \alpha_{m} |m \rangle \, . 
    \label{eq:alfam}
\end{equation}
Applying eq. (\ref{eq:alg1}) to $|m \rangle$ we have  
\begin{equation}
    J_{0} (A^{\dagger} |m \rangle) = A^{\dagger} f(J_{0}) |m \rangle = 
    f(\alpha_{m}) (A^{\dagger} |m \rangle ) \, .
    \label{eq:j+}
\end{equation} 
Thus, we see that $A^{\dagger} |m \rangle $ is a $J_{0}$ 
eigenvector with eigenvalue $f(\alpha_{m})$.  
Starting from $|0 \rangle$ and applying $A^{\dagger}$ successively to 
$|0 \rangle$, we create different states with $J_{0}$ eigenvalue 
given by 
\begin{equation}
    J_{0} \left( (A^{\dagger})^m |0 \rangle \right) = 
    f^m (\alpha_{0}) \left( (A^{\dagger})^m |0 \rangle \right) \, ,
    \label{eq:j+m}
\end{equation}
where $f^m (\alpha_{0})$ denotes the $m$-th iterate of $f$.  Since 
the application of $A^{\dagger}$ creates a new vector, respective 
$J_{0}$ eigenvalue of which has iterations of $\alpha_{0}$ through $f$ 
augmented by one unit, it is 
convenient to define the new vectors $(A^{\dagger})^m |0 \rangle$ as 
proportional to $|m \rangle$ and we then call $A^{\dagger}$ a raising 
operator.  Note that  
\begin{equation}
\alpha_m = f^m(\alpha_0) = f(\alpha_{m-1}) \, ,
    \label{eq:alfam3}
\end{equation}
where $m$ denotes the number of iterations of $\alpha_{0}$ 
through $f$.  

Following the same procedure for $A$,  applying eq. (\ref{eq:alg2}) 
to $|m+1 \rangle$, we have 
\begin{equation}
    A \, J_{0} |m+1 \rangle = f(J_{0}) \left( A |m+1 \rangle \right) =
    \alpha_{m+1} \left( A |m+1 \rangle \right) \, ,
    \label{eq:alfam2}
\end{equation}
showing that $A \, |m+1 \rangle$ is also a $J_{0}$ eigenvector with 
eigenvalue $\alpha_{m}$.  Then, $A \, |m+1 \rangle $ is proportional 
to $|m \rangle$, being $A$ a lowering operator.  

Since we consider 
$\alpha_{0}$ the lowest $J_{0}$ eigenvalue, we require that
\begin{equation}
A \, |0\rangle = 0 .
\label{eq:vacuum}
\end{equation}
As shown in \cite{algebra1}, depending on function $f$ 
and its initial value $\alpha_{0}$, 
$J_{0}$ eigenvalue of state $|m+1 \rangle$ may be lower than that 
of state $|m \rangle$. Then, as shown in \cite{jpa}, 
given an arbitrary analytical function $f$ 
(and its associated algebra in eqs. (\ref{eq:alg1}-\ref{eq:alg3})) 
in order to satisfy eq. (\ref{eq:vacuum}),   
the allowed values of $\alpha_{0}$ are chosen in such a way that the  
iterations $f^m (\alpha_{0})$ ($m \geq 1$) are 
always bigger than $\alpha_{0}$; in other words, eq. (\ref{eq:vacuum})
must be checked for every function $f$, giving consistent vacua
for specific values of $\alpha_0$.

As proven in \cite{jpa}, under the hypothesis stated previously
\footnote{$J_0$ is Hermitian and the existence of a vacuum state.},
for a general function $f$ we obtain
\begin{eqnarray}
J_{0} \, |m\rangle &=& f^{m}(\alpha_0) \, |m\rangle , \; \; \; m = 0,1,2, 
\cdots \; , 
\label{eq:b1} \\
A^{\dagger} \, |m-1\rangle &=& N_{m-1} \, |m\rangle , 
\label{eq:b2} \\
A \, |m\rangle &=& N_{m-1} \, |m-1\rangle ,
\label{eq:b3}
\end{eqnarray}
where $N_{m-1}^2 = f^{m}(\alpha_0)-\alpha_0$. Note that
for each function $f(x)$ the representations are constructed
by the analysis of the above equations as done in
\cite{jpa} for the linear and quadratic $f(x)$.

As shown in \cite{jpa}, where the representation
theory was constructed in detail for the linear and quadratic
functions $f(x)$, the essential tool in order to construct
representations of the algebra in (\ref{eq:alg1}-\ref{eq:alg3})
for a general analytic function $f(x)$ is the analysis of the
stability of the fixed points of $f(x)$ and their composed
functions.

We showed in \cite{comhugo} that there is a class
of one-dimensional quantum systems described by these
generalized Heisenberg algebras. This class is characterized by
those quantum systems having energy eigenvalues written as
\begin{equation}
\epsilon_{n+1} = f(\epsilon_{n}) \, ,
\label{eq:class}
\end{equation}
where $\epsilon_{n+1}$ and $\epsilon_{n}$ are successive energy
levels and $f(x)$ is a different function for each physical
system. This function $f(x)$ is exactly the same function that
appears in the construction of the algebra in eqs. 
(\ref{eq:alg1}-\ref{eq:alg3}). In the algebraic description 
of this class of quantum systems, $J_0$ is the Hamiltonian 
operator of the system, $A^{\dagger}$ and $A$ are the creation 
and annihilation operators. This Hamiltonian and the ladder
operators are related by eq. 
(\ref{eq:casimir}) where $C$ is the Casimir operator of the 
representation associated to the quantum system under
consideration.

\section{Deformed Heisenberg algebra and its physical 
realization} 

In this section we are going to discuss the algebra 
defined by the relations given in eqs. 
(\ref{eq:alg1}-\ref{eq:alg3}) for the linear case, i.e.,
$f(J_0) = r \, J_0 + s$, $s > 0$ $^{\cite{jpa}}$. Furthermore,
we shall also propose a realization, as in the case
of the standard harmonic oscillator, of the ladder 
operators in terms of the physical operators of the
system.

The algebraic relations for the linear case can be rewritten,
after the rescaling
$J_0 \rightarrow s J_0$, $A \rightarrow A/\sqrt{s}$ and
$A^{\dagger} \rightarrow A^{\dagger}/\sqrt{s}$, as
\begin{eqnarray}
\left[ J_{0},A^{\dagger} \right]_{r} &=&  \, A^{\dagger} , 
\label{eq:coml1} \\
\left[ J_{0},A \right]_{r^{-1}} &=& -\frac{1}{r} A , 
\label{eq:coml2} \\
\left[ A^{\dagger},A \right] &=& (1-r) \, J_{0}-1 ,
\label{eq:coml3}
\end{eqnarray}
where $\left[ a,b \right]_{r} \equiv a \, b - r \, b \, a$ is the
$r$-deformed commutation of two operators $a$ and $b$.

It is very simple to realize that, for $r=1$ 
the above algebra is the Heisenberg algebra. 
In this case the Casimir operator given in 
eq. (\ref{eq:casimir}) is
null. Then, for general $r$ the algebra defined in 
eqs. (\ref{eq:coml1}-\ref{eq:coml3})
is a one-parameter deformed Heisenberg algebra and 
generally speaking the algebra
given in eqs. (\ref{eq:alg1}-\ref{eq:alg3}) is a generalization of the
Heisenberg algebra. 

It is easy to see for the general linear case that
\begin{eqnarray}
f^m(\alpha_0) &=& r^m \, \alpha_0 + r^{m-1}+r^{m-2}+
\cdots +1 
\label{eq:gauss1}\\
&=& r^m \alpha_0 +  \frac{r^m -1}{r-1} ,  \nonumber
\end{eqnarray}
thus,
\begin{equation}
N_{m-1}^2 = f^m(\alpha_0)-\alpha_0 = \left[ m \right]_r \, N_{0}^2
\label{eq:gauss2}
\end{equation}
where $\left[ m \right]_r \equiv (r^m -1)/(r-1)$ is the Gauss
number of $m$ and $N_0^2 = \alpha_0 \, (r-1)+1$.

For infinite-dimensional solutions we must solve the following 
set of equations:
\begin{equation}
N_m^2 \; > \; 0 \, , \; \; \; \forall m, \; m = 0,1,2, \cdots \;\; . 
\label{eq:nm2}
\end{equation}
Apart from the Heisenberg algebra given by $r=1$, the solutions
are $^{\cite{jpa}}$
\begin{eqnarray}
\mbox{type} \; \mbox{I (unstable fixed point)} &:& \; r > 1 \;\; 
\mbox{and} \;\; \alpha_0 > 
\frac{1}{1-r} \;\;\;\; 
\mbox{or} 
\label{eq:tipo}\\
\mbox{type} \; \mbox{II   (stable fixed point)    } &:& \; 
-1 < r < 1 \;\; \mbox{and} \;\; \alpha_0 < 
\frac{1}{1-r} \;\; ,
\nonumber
\end{eqnarray}
with matrix representations
\begin{equation}
    J_{0}= \left(  
    \begin{array}{ccccc}
        \alpha_{0} & 0 & 0 & 0 & \ldots  \\
        0 & \alpha_{1} & 0 & 0 & \ldots  \\
        0 & 0 & \alpha_{2} & 0 & \ldots  \\
        0 & 0 & 0 & \alpha_{3} & \ldots  \\
        \vdots & \vdots & \vdots & \vdots & \ddots 
    \end{array}
    \right)  , \hspace{0.1cm}
       A^{\dagger}= \left(  
    \begin{array}{ccccc}
        0 & 0 & 0 & 0 & \ldots  \\
        N_{0} & 0 & 0 & 0 & \ldots  \\
        0 & N_{1} & 0 & 0 & \ldots  \\
        0 & 0 & N_{2} & 0 & \ldots  \\
        \vdots & \vdots & \vdots & \vdots & \ddots 
    \end{array}
    \right) , \hspace{0.1cm}
    A = (A^{\dagger})^{\dag} \hspace{0.1cm} .
    \label{eq:matriz}
\end{equation}
Note that for type I solutions the eigenvalues of $J_0$ , as can be 
easily computed
from eqs. (\ref{eq:b1})  and (\ref{eq:alfam3}), go to infinite as 
we consider eigenvectors 
$|m\rangle$ 
with increasing value of $m$. Instead, for type II solutions the
eigenvalues go to the value $1/(1-r)$, the fixed point of $f$,
as the state $|m\rangle$ increase.

It is easy to see that there is a direct relation between
the linear Heisenberg algebra given in 
eqs. (\ref{eq:coml1}-\ref{eq:coml3}) and the standard
$q$-oscillators. In fact, defining $^{\cite{jpa}}$
\begin{eqnarray}
J_0 & = & q^{2 N} \, \alpha_0 +  \left[ N \right]_{q^2} \;\; , 
\label{eq:qosc1} \\
\frac{A^{\dagger}}{N_0} & = & a^{\dagger} \,  q^{N/2} \;\; ,
\label{eq:qosc2}\\
\frac{A}{N_0} & = & q^{N/2} \, a \;\; , 
\label{eq:qosc3}
\end{eqnarray}
we see that $a$, $a^{\dagger}$ and $N$ satisfy the usual 
$q$-oscillator relations \cite{qosc}
\begin{eqnarray}
&&a \, a^{\dagger} - q \, a^{\dagger} \, a = q^{-N} \;\; , \;\;
a \, a^{\dagger} - q^{-1} \, a^{\dagger} \, a = q^{N} \;\; , 
\label{eq:qosc4}\\
&&\left[ N,a \right] = -a \;\; , \;\;
\left[ N,a^{\dagger} \right] = a^{\dagger} \;\; . \nonumber 
\end{eqnarray}
Note that Heisenberg algebra is obtained from 
(\ref{eq:qosc1}-\ref{eq:qosc3}) for
$q \rightarrow 1$ and $\alpha_0 = 0$.

The next step we have to take is to realize the operators $A$, 
$A^{\dagger}$ and $J_0$ in terms of physical
operators as in the case of the one-dimensional
harmonic oscillator, and as it was done in
\cite{comhugo} and in \cite{campos} for the
square-well potential. To do this, we briefly 
review the formalism of non-commutative differential 
and integral calculus on a one-dimensional lattice developed
in \cite{dimakis1} and \cite{dimakis2}. Let us consider a  
one-dimensional lattice in a momentum 
space where the 
momenta are allowed only to take discrete values, say $p_{0}$, 
$p_{0}+a$, $p_{0}+2a$, $p_{0}+3a$ etc, with $a>0$.

The non-commutative differential calculus is based on the 
expression $^{\cite{dimakis1}, \cite{dimakis2}}$ 
\begin{equation}
    [p,dp] = dp \, a
    \label{eq:noncom1} \, ,
\end{equation}
implying that 
\begin{equation}
    f(p) \, dg(p) = dg(p) \, f(p+a) \, ,
    \label{eq:noncom2}
\end{equation}
for all functions $f$ and $g$. We introduce partial 
derivatives as 
\begin{equation}
    d \, f(p) = dp \, (\partial_{p} \, f) \, (p) = 
    (\bar{\partial}_{p} \, f) \, (p) \, dp \, ,
    \label{partial}
\end{equation}    
where the left and right discrete  derivatives are given by 
\begin{eqnarray}
    (\partial_{p} \, f) \, (p) & = & \frac{1}{a} \, [f(p+a) - f(p)] \, ,
    \label{eq:partialleft}  \\
   (\bar{\partial}_{p} \, f) \, (p)  & = & \frac{1}{a} \, [f(p) - f(p-a)] \, , 
    \label{eq:partialright}
\end{eqnarray}
that are the two possible definitions of derivatives on a lattice.
The Leibniz rule for the left discrete derivative can be written as,
\begin{equation}
    (\partial_{p} \, fg) \, (p) = (\partial_{p}f) \, (p) g \, (p) + 
    f(p+a)(\partial_{p} g) \, (p) \, ,
    \label{eq:leibniz}
\end{equation}
with a similar formula for the right derivative $^{\cite{dimakis1}}$.

Let us now introduce the momentum shift operators 
\begin{eqnarray}
    T  & = & 1 + a \, \partial_{p}
    \label{eq:a}  \\
    \bar{T} & = & 1 - a \, \bar{\partial}_{p} \, ,
    \label{eq:abarra}
\end{eqnarray}
that shift the momentum value by $a$
\begin{eqnarray}
    (Tf) \, (p) & = & f(p+a)
    \label{eq:af}  \\
    (\bar{T}f) \, (p) & = & f(p-a)
    \label{eq:abarraf}
\end{eqnarray}
and satisfies 
\begin{equation}
    T \, \bar{T} = \bar{T} T = \hat{1} \, ,
    \label{eq:aabarra}
\end{equation}
where $\hat{1}$ means the identity on the algebra of functions of $p$.  

Introducing the momentum operator $P$ $^{\cite{dimakis1}}$
\begin{equation}
    (Pf) \, (p) = p \, f(p) \, ,
    \label{eq:momentum}
\end{equation}
we have 
\begin{eqnarray}
    T P & = & (P+a)T
    \label{eq:ap}  \\
    \bar{T} P & = & (P-a) \bar{T} \, \, .
    \label{eq:abarrap}
\end{eqnarray}

Integrals can also be defined in this formalism. It is shown
in ref. \cite{dimakis1} that the property of an indefinite
integral 
\begin{equation}
\int df = f + \mbox{periodic function in} \, \, a \,\, ,
    \label{eq:intindef}
\end{equation}
suffices to calculate the indefinite integral of an
arbitrary one form. It can be shown that $^{\cite{dimakis1}}$
for an arbitrary function $f$
\begin{equation}
\int d\bar{p} \, f(\bar{p}) = \left\{
\begin{array}{lll}
a \sum_{k=1}^{[p/a]} f(p-k a) \, , & \mbox{if $p \geq a$} \\
0 \, \, ,  & \mbox{if $0 \leq p < a$} \\
-a \sum^{-[p/a]-1}_{k=0} f(p+k a) \, ,  & \mbox{if $p < 0$}
\end{array}
\right.
\label{eq:intresult}
\end{equation}
where $[p/a]$ is by definition the highest integer $\leq p/a $.

All equalities involving indefinite integrals are understood
modulo the addition of an arbitrary function periodic in $a$.
The corresponding definite integral is well-defined when the 
length of the interval is multiple of $a$. Consider the integral
of a function $f$ from $p_d$ to $p_u$ ($p_u = p_d+M a$, where
$M$ is a positive integer) as
\begin{equation}
\int_{p_d}^{p_u} dp f(p) = a \sum_{k=0}^{M} f(p_d+k a).
    \label{eq:intdef}
\end{equation}
Using eq. (\ref{eq:intdef}), 
an inner product of two (complex) functions 
$f$ and $g$ can be defined as 
\begin{equation}
    \langle f \, , \, g \rangle = \int_{p_{d}}^{p_{u}} dp \, f(p)^{*} \, g(p) \, , 
    \label{eq:inner}
\end{equation}
where $^{*}$ indicates the complex conjugation of the function $f$.  
The norm $\langle f \, , \, f \rangle \geq 0$ is
zero only when $f$ is identically null.  The set of equivalence classes  
\footnote{Two functions are in the same equivalence class if their
values coincide on all lattice sites.}
of normalizable functions $f$ ($\langle f \, , \, f \rangle $ is finite) 
is a Hilbert space.  
It can be shown that $^{\cite{dimakis1}}$
\begin{equation}
    \langle f, T g \rangle = \langle \bar{T} f, g \rangle \, ,
    \label{eq:inner2}
\end{equation}
so that  
\begin{equation}
    \bar{T} = T^\dagger \, \, ,
    \label{eq:adjoint}
\end{equation}
where $T^\dagger$ is the adjoint operator of $T$.  Eqs. (\ref{eq:aabarra}) and 
(\ref{eq:adjoint}) show that $T$ is a unitary operator. Moreover, it is easy
to see that $P$ defined in eq. (\ref{eq:momentum}) is an Hermitian operator
and from (\ref{eq:adjoint}) one has
\begin{equation}
(i \partial_p)^{\dagger} = i \bar{\partial}_p  \, \, .
\label{eq:hermopp}
\end{equation} 

Now, we go back to the realization of the deformed
Heisenberg algebra eqs. 
(\ref{eq:coml1}-\ref{eq:coml3}) in terms of 
physical operators. We can associate to the one-parameter
deformed Heisenberg algebra in eqs. 
(\ref{eq:coml1}-\ref{eq:coml3}) the one-dimensional 
lattice we have just presented.

Observe that we can write $J_0$ in this case as
\begin{equation}
J_0 = q^{2 P/a} \, \alpha_0 +  \left[ P/a \right]_{q^2} \, \, ,
\label{eq:defj0}
\end{equation} 
where $P$ is given in eq. (\ref{eq:momentum}) and
its application to the vector states $|m\rangle$
appearing in (\ref{eq:b1}-\ref{eq:b3}) gives $^{\cite{comhugo}}$
\begin{equation}
P \, |m\rangle = m \, a \, |m\rangle \,\, , m=0,1, 
\cdots  \, \, ,
\label{eq:aplicmom}
\end{equation}
where we can write $N=P/a$ with $N |m\rangle = m |m\rangle$.
Moreover,
\begin{equation}
\bar{T} \, |m\rangle = |m+1\rangle \,\, , m=0,1, 
\cdots  \, \, ,
\label{eq:aplictbar}
\end{equation}
where $\bar{T}$ and $T=\bar{T}^{\dagger}$ are defined
in eqs. (\ref{eq:a}-\ref{eq:aabarra}).

With the definition of $J_0$ given in eq. (\ref{eq:defj0})
we see that $\alpha_n$ given in eq.                
(\ref{eq:alfam3}) is the $J_0$ eigenvalue of
state $|n\rangle$ as we wanted. Let us now define
\begin{eqnarray}
A^{\dagger}  &=& S(P) \, \bar{T} \,\,  ,
\label{eq:real1} \\
A &=& T \, S(P) \,\,  ,
\label{eq:real2} 
\end{eqnarray}
where,
\begin{equation}
S(P)^2 = J_0 - \alpha_0 \, \, ,
\label{eq:defS}
\end{equation}
where $\alpha_0$,
defined in eq. (\ref{eq:alfa0}), is the lowest $J_0$ eigenvalue.

Yet, note that
eqs. (\ref{eq:ap}-\ref{eq:abarrap}) can also be
rewritten as
\begin{eqnarray}
    T N & = & (N+1)T
    \label{eq:comutTN1}  \\
    \bar{T} N & = & (N-1) \bar{T} \, \, .
    \label{eq:comutTN2}
\end{eqnarray}

It is easy to realize that $A$, $A^{\dagger}$ and
$J_0$ defined in eqs. (\ref{eq:defj0}, 
\ref{eq:real1}-\ref{eq:defS}) satisfy the one-parameter deformed
algebra given in eqs. (\ref{eq:coml1}-\ref{eq:coml3}).
Consider firstly the relation between $J_0$ and $A^{\dagger}$,
\begin{equation}
J_0 \, A^{\dagger} = \alpha_N \,\,\, S(P) \, \bar{T} =
A^{\dagger} \, \alpha_{N+1}  \, \, ,
\label{eq:prova1}
\end{equation}
where $\alpha_N$ is $\alpha_m = f^m(\alpha_0)$ in eq. (\ref{eq:gauss1})
with the operator $N$ in place of variable $m$.                                  
In (\ref{eq:prova1}) we have used the realizations in the first
equality of the above equation and in the second one
we have used
eq. (\ref{eq:comutTN2}). But, from eq. (\ref{eq:alfam3})
$\alpha_{N+1}= f(\alpha_N)=f(J_0)$ thus we obtain,
\begin{equation}
J_0 \, A^{\dagger} = 
A^{\dagger} \, f(J_0)  \, \, ,
\label{eq:prova2}
\end{equation}
that is, eq. (\ref{eq:alg1}) for $f(x)$ linear.
Eq. (\ref{eq:alg2}) is the
Hermitian conjugate of eq. (\ref{eq:alg1}), then
its proof using eq. (\ref{eq:real2}) and (\ref{eq:defj0})
is similar to the previous one. Now, using
\begin{eqnarray}
    A^{\dagger} \, A & = & S(P)^2 = J_0 - \alpha_0 \,\, ,
    \label{eq:prova3}  \\
    A \, A^{\dagger} & = & T \, S(P)^2 \bar{T} = f(J_0) - \alpha_0
 \, \, ,
    \label{eq:prova4}
\end{eqnarray}
for linear $f(x)$, that has the property given in            
eq. (\ref{eq:alfam3}), we get
eq. (\ref{eq:alg3}) for $f(x)$ linear and the proof is 
complete.

Note that the realization we have found, as shown in 
eqs. (\ref{eq:real1}, \ref{eq:real2} and 
\ref{eq:defj0}), is qualitatively different from the 
realization of the standard harmonic oscillator.
This shows that the realization of the ladder operators
for the harmonic oscillator means that the deformation
parameter $r=1$ is a special case.

\section{A deformed quantum field theory}

We are going to discuss in this section a QFT
having as excitations objects
described by the one-parameter deformed
algebra given in eqs. (\ref{eq:coml1}-\ref{eq:coml3}). 
In this QFT the mass spectrum consists of only one
particle with mass $m$. In this case the energy of $n$ particles
is not equal to $n$ times the energy of one particle and therefore
the energy does not obey the additivity rule. 
This non-additivity comes from the fact 
that $q$-oscillators
approximately describe correlated fermion pairs in many body systems 
$^{\cite{bonatsos}}$. The advantage of this construction is that, 
being a deformation,  we
can make a contact with the well-known non-deformed 
model in all steps of the computation by taking the
deformation parameter going to one. 

In the momentum space appropriated to the realization of
the deformed Heisenberg algebra we discussed, besides
the operator $P$ defined in eq. (\ref{eq:momentum}),
one can define two self adjoint operators as
\begin{eqnarray}
    \chi & \equiv & - i \left( S(P)(1-a \bar{\partial}_p )- 
(1+a \partial_p )S(P) \right) =
- i(A - A^{\dagger})  \, \, ,
    \label{eq:cord1}  \\
     Q & \equiv &  S(P)(1-a \bar{\partial}_p )+ 
(1+a \partial_p )S(P) =
A + A^{\dagger}  \, \, ,
    \label{eq:cord2}  
\end{eqnarray}
where $\partial_p$ and $\bar{\partial}_p$ are the left and right 
discrete derivatives defined in eqs. (\ref{eq:partialleft}, 
\ref{eq:partialright}).

It can be checked that operators $P$, $\chi$ and $Q$ generate
the following algebra on the momentum lattice:
\begin{eqnarray}
\left[ \chi,P \right] &=&  i a Q  ,
\label{eq:fecho1} \\
\left[ P,Q \right] &=&  i a \chi , 
\label{eq:fecho2} \\
\left[ \chi,Q \right] &=& 2 i S(P) \left( S(P+a)-S(P-a) \right) .
\label{eq:fecho3}
\end{eqnarray}

To construct a QFT based on these operators 
let us now introduce a three-dimensional discrete $\vec{k}$-space,
\begin{equation}
k_i = \frac{2 \pi l_i}{L_i} ,\, \,\,\,  i=1,2,3 \,\,\,\,\,\,\, ,
\label{eq:kspace}
\end{equation}
with $l_i= 0,\pm 1,\pm 2, \cdots $ and $L_i$ being the lengths of the 
three sides of a rectangular box $\Omega$. We introduce for
each point of this $\vec{k}$-space an independent copy of the $r$-deformed
harmonic oscillator constructed in the last two previous
sections so that the deformed operators commute for different
three-dimensional lattice points. 
We also introduce an independent copy of the one-dimensional
momentum lattice defined in the previous section  
for each point of this $\vec{k}$-lattice  
so that $P_{\vec{k}}^{\dagger} = P_{\vec{k}}$ and 
$T_{\vec{k}}$, $\bar{T}_{\vec{k}}$ and $S_{\vec{k}}$ 
are defined by means of the previous definitions,
eqs. (\ref{eq:a}-\ref{eq:abarra} and \ref{eq:defS}),
through the substitution $P \rightarrow P_{\vec{k}}$.

It is not difficult to realize that
\begin{eqnarray}
    A^{\dagger}_{\vec{k}} & = &  S_{\vec{k}} \, \bar{T}_{\vec{k}} \, ,
    \label{eq:j+k}  \\
     A_{\vec{k}} & = & T_{\vec{k}} \, S_{\vec{k}} \, ,    
    \label{eq:j-k}  \\
    J_0(\vec{k}) &=& q^{2 P_{\vec{k}}/a} \, \alpha_0 +  
\left[ P_{\vec{k}}/a \right]_{q^2} \, \, ,
    \label{eq:j0k}
\end{eqnarray}
satisfy the algebra in eqs. (\ref{eq:coml1}-\ref{eq:coml3})
for each point of this $\vec{k}$-lattice and the operators
$A^{\dagger}_{\vec{k}}$, $A_{\vec{k}}$ and $J_{0}(\vec{k})$ 
commute among them for different points of this
$\vec{k}$-lattice.

Now, we define operators $\chi$ and $Q$ for
each point of the three-dimensional lattice as
\begin{eqnarray}
    \chi_{\vec{k}} & \equiv & -i(T_{-\vec{k}} \, S_{-\vec{k}} - 
S_{\vec{k}} \, \bar{T}_{\vec{k}} ) =
- i ( A_{-\vec{k}} - A^{\dagger}_{\vec{k}})   \, \, ,
    \label{eq:cord5}  \\
     Q_{\vec{k}} & \equiv &  T_{\vec{k}} \, S_{\vec{k}} + 
S_{-\vec{k}} \, \bar{T}_{-\vec{k}} =
A_{\vec{k}} + A^{\dagger}_{-\vec{k}}  \, \, ,
    \label{eq:cord6}  
\end{eqnarray}
such that $\chi_{\vec{k}}^{\dagger}= \chi_{-\vec{k}}$ and 
$Q_{\vec{k}}^{\dagger}= Q_{-\vec{k}}$, exactly as it happens
in the construction of a spin-$0$ field for the spin-$0$ quantum 
field theory $^{\cite{tdlee}}$. 

By means of $\chi_{\vec{k}}$ and  $Q_{\vec{k}}$ we define 
two fields
$\phi(\vec{r},t)$ and $\Pi(\vec{r},t)$ as
\begin{eqnarray}
\phi(\vec{r},t) &=&  \sum_{\vec{k}} 
\frac{1}{\sqrt{2\Omega\omega(\vec{k})}}  \left( 
A^{\dagger}_{\vec{k}} \, e^{-i \vec{k}.\vec{r}} + 
A_{\vec{k}} \, e^{i \vec{k}.\vec{r}} \right)  ,
\label{eq:defcampo1} \\
\Pi(\vec{r},t) &=&  \sum_{\vec{k}} 
\frac{i\omega(\vec{k})}{\sqrt{2\Omega\omega(\vec{k})}} \left( 
A^{\dagger}_{\vec{k}} \, e^{-i \vec{k}.\vec{r}}- 
A_{\vec{k}} \, e^{i \vec{k}.\vec{r}} \right) , 
\label{eq:defcampo2} 
\end{eqnarray}
where $\omega(\vec{k})= \sqrt{\vec{k}^2+m^2}$, $m$ is a real parameter
and $\Omega$ is the volume of a rectangular box. Another momentum-type
field  $\wp(\vec{r},t)$ can be defined as 
\begin{equation}
\wp(\vec{r},t) = \sum_{\vec{k}} \sqrt{\frac{\omega(\vec{k})}{2\Omega}}  
\, \, \, S_{\vec{k}}  \, e^{i \vec{k}.\vec{r}}  \, .
\label{eq:defcampo3}  
\end{equation}

By a straightforward calculation, we can show that the 
Hamiltonian
\begin{eqnarray}
H =  &\int& d^3 r \left(  \Pi(\vec{r},t)^2
 +u \, | \wp(\vec{r},t)|^2 + 
   \phi(\vec{r},t)  (-{\vec{\nabla}}^2+m^2)  
\phi(\vec{r},t)   \right)  \,\, ,
\label{eq:defhamilt}
\end{eqnarray}
where $u$ is an arbitrary number, can be written as
\begin{eqnarray}
H &=& \frac{1}{2} \sum_{\vec{k}} \omega(\vec{k}) 
\left( A^{\dagger}_{\vec{k}} A_{\vec{k}}+A_{\vec{k}} 
A^{\dagger}_{\vec{k}}+u \, S_{\vec{k}}(N)^2 \right) \nonumber \\
&=&\frac{1}{2}\sum_{\vec{k}} \omega(\vec{k}) \left( S_{\vec{k}}(N+1)^2 +
(1+u) \, S_{\vec{k}}(N)^2 \right)   \,\,  ,
\label{eq:resulthamilt}
\end{eqnarray}
where
\begin{equation}
S_{\vec{k}}(N)^2 = q^{2 N_{\vec{k}}} \, \alpha_0 + 
\left[ N_{\vec{k}} \right]_{q^2} - \alpha_0 \, \, .
\label{eq:defs2}
\end{equation}
In order that the energy of the vacuum state becomes zero
we replace $H$ in eq. (\ref{eq:resulthamilt}) by
\begin{equation}
H = \frac{1}{2}\sum_{\vec{k}} \omega(\vec{k}) \left( S_{\vec{k}}(N+1)^2 +
(1+u) \, S_{\vec{k}}(N)^2 - (q^2-1)\alpha_0 -1 \right) \,\, .
\label{eq:resulthamilt1}
\end{equation}
Note that in the limit $q \rightarrow 1$ ($\alpha_0 \rightarrow 0$),
the above Hamiltonian is proportional to the number operator.
Furthermore, as can be seen from eqs.(\ref{eq:defs2}) and 
(\ref{eq:resulthamilt1}) the energy of this system is non-additive.
The non-additivity of a free system is not new in field theory,
for instance the energy of non-topological solitons has this property
(see chapter 7 in \cite{tdlee}). In the present case, this 
non-additivity of the energy comes from the fact that $q$-oscillators 
approximately describe correlated fermion pairs in many-body systems
$^{\cite{bonatsos}}$. 

The eigenvectors of $H$ form a complete set and span
the Hilbert space of this system, they are
\begin{equation} 
|0 \rangle, \,\, A^{\dagger}_{\vec{k}} |0 \rangle, \,\,
A^{\dagger}_{\vec{k}} A^{\dagger}_{\vec{k}'} |0 \rangle \,\,
\mbox{for} \,\, \vec{k}\not= \vec{k}', \,\,
(A^{\dagger}_{\vec{k}})^2 |0 \rangle, \,\, \cdots \,\, ,
\label{eq:hilbert} 
\end{equation}
where the state $|0\rangle$ satisfies as usual 
$A_{\vec{k}} |0\rangle =0$ (see eq. 
(\ref{eq:vacuum})) for all $\vec{k}$ and $A_{\vec{k}}$, 
$A^{\dagger}_{\vec{k}}$ for each $\vec{k}$
satisfy the $q$-deformed Heisenberg algebra 
eqs. (\ref{eq:coml1}-\ref{eq:coml3}).

The time evolution of the fields can be studied by
means of Heisenberg's equation for $A^{\dagger}_{\vec{k}}$,
$A_{\vec{k}}$ and $S_{\vec{k}}$. Define
\begin{equation}
E(N_{\vec{k}}) \equiv J_0(\vec{k}) = q^{2 N_{\vec{k}}} \, 
\alpha_0 +  \left[ N_{\vec{k}} \right]_{q^2} 
\label{eq:defen}  
\end{equation}
and
\begin{equation}
h(N_{\vec{k}})  \equiv  \frac{1}{2} (1+u+r) \left(
 E(N_{\vec{k}}+1) - E(N_{\vec{k}}) \right) \,\, .
\label{eq:defh}  
\end{equation}
Thus, using eqs. (\ref{eq:resulthamilt} or \ref{eq:resulthamilt1})
and (\ref{eq:coml1}-\ref{eq:coml3}) we obtain
\begin{equation}
\left[ H, A^{\dagger}_{\vec{k}} \right] = 
\omega(\vec{k}) \, A^{\dagger}_{\vec{k}} \,\, h(N_{\vec{k}})   \,\, .
\label{eq:comutheis1}  
\end{equation}
We can solve Heisenberg's equation for the $q$-deformed case
obtaining
\begin{equation}
A^{\dagger}_{\vec{k}}(t) =  A^{\dagger}_{\vec{k}}(0) \,\,
e^{i \omega(\vec{k}) \, h(N_{\vec{k}}) \, t}   \,\, .
\label{eq:solveheis1}  
\end{equation}
Note that for $q \rightarrow 1$ and $u=0$ we have 
$h(N_{\vec{k}}) \rightarrow 1$ and eq. (\ref{eq:solveheis1})
gives the correct result for this undeformed case.
It is not difficult to realize, using 
eqs. (\ref{eq:coml1}-\ref{eq:coml3}), that for this 
linear case we have $(f(J_0)-J_0)\,A = q^{-2}\, A \, 
(f(J_0)-J_0)$. Taking the Hermitean conjugate of
eq. (\ref{eq:solveheis1}) and using the result just mentioned
we have
\begin{equation}
A_{\vec{k}}(t) =  A_{\vec{k}}(0) \,\,
e^{-i \, q^{-2} \, \omega(\vec{k}) \, h(N_{\vec{k}}) \, t}   \,\, ,
\label{eq:solveheis2}  
\end{equation}
giving also the correct undeformed limit. Furthermore, we easily see 
that operators $P_{\vec{k}}$ and $S_{\vec{k}}$ are 
time-independent. We emphasize that the extra term, $h(N_{\vec{k}})$,
in the exponentials depends on the number operator, being this the
main difference from the undeformed case. The Fourier
transformation eq. (\ref{eq:defcampo1}) can then be written as
\begin{equation}
\phi(\vec{r},t) = \alpha(\vec{r},t) + \alpha(\vec{r},t)^{\dagger}
 \,\, ,
\label{eq:defcampotempo}  
\end{equation}
where
\begin{equation}
\alpha(\vec{r},t) =  \sum_{\vec{k}} 
\frac{1}{\sqrt{2\Omega\omega(\vec{k})}} \,\,   
A_{\vec{k}} \, e^{i \vec{k}.\vec{r}-i q^{-2}\omega(\vec{k}) \, 
h(N_{\vec{k}}) \, t}  \,\, ,
\label{eq:defcampotempo1}  
\end{equation}
$A_{\vec{k}}$ in eq. (\ref{eq:defcampotempo1}) is  
time-independent and $\alpha(\vec{r},t)^{\dagger}$
is the Hermitean conjugate of $\alpha(\vec{r},t)$.

The Feynman propagator $D^{N}_F(x_1,x_2)$ defined, as usual, 
as the Dyson-Wick contraction between 
\footnote{$x_i \equiv (\vec{r_i},t_i)$}
$\phi(x_1)$ and $\phi(x_2)$,
can be computed using eqs. (\ref{eq:coml1}-\ref{eq:coml3} and
\ref{eq:defcampotempo}-\ref{eq:defcampotempo1})
being 
\begin{equation}
D^{N}_F(x_1,x_2) = \sum_{\vec{k}} \frac{e^{i\vec{k} . \Delta 
\vec{r}_{12}}}{2\Omega\omega(\vec{k})}
\left( S_{\vec{k}}(N+1)^2 \, e^{\mp i \omega(\vec{k})\, h(N_{\vec{k}})\,
\Delta t_{12}} - S_{\vec{k}}(N)^2 \, e^{\mp i \omega(\vec{k})\, 
h(N_{\vec{k}}-1)\, \Delta t_{12}} 
\right) \, ,
\label{eq:propsum}
\end{equation}
where $\Delta t_{12}=t_1-t_2$, $\Delta \vec{r}_{12}=
\vec{r}_1-\vec{r}_2$, the minus sign in the exponent
holds when $t_1 > t_2$ and the positive sign when
$t_2 > t_1$. Note that when $q \rightarrow 1$, 
$h(N_{\vec{k}}) \rightarrow 1$ and $S_{\vec{k}}(N+1)^2 -
S_{\vec{k}}(N)^2 \rightarrow 1$, the 
standard result for the propagator is recovered. It is also
simple to obtain the following integral representation
for the Feynman propagator
\begin{equation}
D^{N}_F(x) = \frac{-i}{(2\pi)^4} \int \frac{S_{\vec{k}}(N+1)^2
e^{i\vec{k}.\vec{r}-ik_0\, h(N_{\vec{k}}) \, t}\,\, d^4 k}{k^2+m^2} -
(N \rightarrow N-1) \, ,
\label{eq:propint}
\end{equation}
where in the second part of the right hand side of the above equation 
we have the first part with $N \rightarrow N-1$. Again, note that
the $q \rightarrow 1$ limit of the integral representation of the
above Feynman propagator gives the usual propagator. We point out
that this propagator is not a simple c-number since it depends on
the number operator $N$.

We shall now discuss the first order scattering process 
$1+2 \rightarrow 1^{'}+2^{'}$ where the initial state is
\begin{equation}
|1,2 \rangle \equiv A^{\dagger}_{p_1} \, A^{\dagger}_{p_2} |0\rangle
\,\, , 
\label{eq:estinic}
\end{equation} 
and the final state is
\begin{equation}
|1^{'},2^{'} \rangle \equiv A^{\dagger}_{p_1^{'}} \, 
A^{\dagger}_{p_2^{'}} |0\rangle
\,\, , 
\label{eq:estfin}
\end{equation}
where $A_{p_i}$ and $A^{\dagger}_{p_i}$ satisfy the algebraic 
relations in eqs. (\ref{eq:coml1}-\ref{eq:coml3}).These
particles are supposed to be described by the Hamiltonian
given in eq. (\ref{eq:defhamilt}) with an interaction
given by $\lambda \int :\phi(\vec{r},t)^4: d^3r$. To the
lowest order of $\lambda$, we have
\begin{equation}
\langle 1^{'},2^{'} | S | 1,2 \rangle = -i \lambda \int 
d^4 x \langle 1^{'},2^{'} | :\phi^4(x): | 1,2 \rangle \,\, .
\label{eq:matrixelem1}
\end{equation}
We use eq. (\ref{eq:defcampotempo}) in eq. (\ref{eq:matrixelem1})
and put it in normal order. Since the exponentials have now number
operators we must take exponentials in it outside 
from the matrix elements obtaining
\begin{eqnarray}
\langle 1^{'},2^{'} | S | 1,2 \rangle = 
\frac{-6i\lambda}{4\Omega^2} \int d^4 x \sum_{\vec{k}_1
\cdots \vec{k}_4} \frac{1}{\sqrt{\omega_{\vec{k}_1}\cdots 
\omega_{\vec{k}_4} }}
\langle 0 | A_{\vec{p'}_1} A_{\vec{p'}_2} A^{\dagger}_{\vec{k}_1}
A^{\dagger}_{\vec{k}_2} A_{\vec{k}_3} A_{\vec{k}_4} 
A^{\dagger}_{\vec{p}_1} A^{\dagger}_{\vec{p}_2} | 0 \rangle 
\nonumber \\
e^{-i(\vec{k}_1+\vec{k}_2-\vec{k}_3-\vec{k}_4).\vec{r}
+ i W(\vec{k}_1,\vec{k}_2,\vec{k}_3,\vec{k}_4)t}  \,\, , \nonumber \\
\label{eq:matrixelem2}
\end{eqnarray}
where
\begin{eqnarray}
W(\vec{k}_1,\vec{k}_2,\vec{k}_3,\vec{k}_4) = \omega(\vec{k}_1) \,
h_1(\vec{k}_1,\vec{k}_2,\vec{k}_3,\vec{k}_4) + 
\omega(\vec{k}_2) \, h_2(\vec{k}_1,\vec{k}_2,\vec{k}_3,\vec{k}_4)-
 \nonumber \\
\omega(\vec{k}_3) \, h_3(\vec{k}_1,\vec{k}_2,\vec{k}_3,\vec{k}_4)-
\omega(\vec{k}_3) \, h_3(\vec{k}_1,\vec{k}_2,\vec{k}_3,\vec{k}_4) \,\, ,
\label{eq:defW}
\end{eqnarray}
and
\begin{eqnarray}
h_1(\vec{k}_1,\vec{k}_2,\vec{k}_3,\vec{k}_4) &=& h \left( 
\delta^3_{\vec{k}_1,\vec{k}_2} - \delta^3_{\vec{k}_1,\vec{k}_3}-
\delta^3_{\vec{k}_1,\vec{k}_4} + \delta^3_{\vec{k}_1,\vec{p}_1} +
\delta^3_{\vec{k}_1,\vec{p}_2} \right) \,\, , \nonumber \\
h_2(\vec{k}_1,\vec{k}_2,\vec{k}_3,\vec{k}_4) &=& h \left( 
-\delta^3_{\vec{k}_2,\vec{k}_3} - \delta^3_{\vec{k}_2,\vec{k}_4}+
\delta^3_{\vec{k}_2,\vec{p}_1} + \delta^3_{\vec{k}_2,\vec{p}_2}
\right) \,\, , \nonumber \\
h_3(\vec{k}_1,\vec{k}_2,\vec{k}_3,\vec{k}_4) &=& h \left(
-1 - \delta^3_{\vec{k}_3,\vec{k}_4} + \delta^3_{\vec{k}_3,\vec{p}_1}
+ \delta^3_{\vec{k}_3,\vec{p}_2} \right) \,\, , \nonumber \\
h_4(\vec{k}_1,\vec{k}_2,\vec{k}_3,\vec{k}_4) &=& h \left(
-1 + \delta^3_{\vec{k}_4,\vec{p}_1} + \delta^3_{\vec{k}_4,\vec{p}_2}
\right) \,\, .
\label{eq:defdosh}
\end{eqnarray}
The matrix elements in eq. (\ref{eq:matrixelem2}) can be handled
using the algebraic relations in eqs. (\ref{eq:coml1}-\ref{eq:coml3}).
At this point the integral in eq. (\ref{eq:matrixelem2}) can be 
computed giving
\begin{eqnarray}
&&\langle 1^{'},2^{'} | S | 1,2 \rangle = \frac{-24(2\pi)^4ih(0)^2
\lambda}{\Omega^2 (1+u+q^2)^4\sqrt{\omega_{\vec{p}_1}\omega_{\vec{p}_2} 
\omega_{\vec{p'}_1}\omega_{\vec{p'}_2} }} \left( h(0)^2 
\delta^4(P_{1,a}+P_{2,a}-P'_{1,a}-P'_{2,a}) + \right. \nonumber \\
&& h(0) h(\delta^3_{\vec{p'}_1,\vec{p'}_2}) \delta^4(P_{1,b}+P_{2,b}-
P'_{1,b}-P'_{2,b})+h(0) h(\delta^3_{\vec{p}_1,\vec{p}_2}) \delta^4
(P_{1,c}+P_{2,c}-P'_{1,c}-P'_{2,c})+ \nonumber \\
&& \left. h(\delta^3_{\vec{p}_1,\vec{p}_2}) h(\delta^3_{\vec{p'}_1,
\vec{p'}_2}) \delta^4(P_{1,d}+P_{2,d}-P'_{1,d}-P'_{2,d}) 
\right) \,\, ,
\label{eq:finalmatrix}
\end{eqnarray}
where
\begin{eqnarray}
P_{1,a} = (\vec{p}_1,\omega_{\vec{p}_1}h(0)),  
P_{2,a} = (\vec{p}_2,\omega_{\vec{p}_2}h(\delta^3_{\vec{p}_1,
\vec{p}_2})), 
P'_{1,a} = (\vec{p'}_1,\omega_{\vec{p'}_1}h(\delta^3_{\vec{p'}_1,
\vec{p'}_2})),  
P'_{2,a} = (\vec{p'}_2,\omega_{\vec{p'}_2}h(0));  \nonumber \\
P_{1,b} = (\vec{p}_1,\omega_{\vec{p}_1}h(0)),  
P_{2,b} = (\vec{p}_2,\omega_{\vec{p}_2}h(\delta^3_{\vec{p}_1,
\vec{p}_2})), 
P'_{1,b} = (\vec{p'}_1,\omega_{\vec{p'}_1}h(0)),  
P'_{2,b} = (\vec{p'}_2,\omega_{\vec{p'}_2}h(\delta^3_{\vec{p'}_1,
\vec{p'}_2}));  \nonumber \\
P_{1,c} = (\vec{p}_1,\omega_{\vec{p}_1}h(\delta^3_{\vec{p}_1,
\vec{p}_2})),  
P_{2,c} = (\vec{p}_2,\omega_{\vec{p}_2}h(0)), 
P'_{1,c} = (\vec{p'}_1,\omega_{\vec{p'}_1}h(\delta^3_{\vec{p'}_1,
\vec{p'}_2})),  
P'_{2,c} = (\vec{p'}_2,\omega_{\vec{p'}_2}h(0));  \nonumber \\
P_{1,d} = (\vec{p}_1,\omega_{\vec{p}_1}h(\delta^3_{\vec{p}_1,
\vec{p}_2})),  
P_{2,d} = (\vec{p}_2,\omega_{\vec{p}_2}h(0)),
P'_{1,d} = (\vec{p'}_1,\omega_{\vec{p'}_1}h(0)),  
P'_{2,d} = (\vec{p'}_2,\omega_{\vec{p'}_2}h(\delta^3_{\vec{p'}_1,
\vec{p'}_2})). \nonumber \\
\label{eq:definePs}
\end{eqnarray}
Note that when $q \rightarrow 1$ we have $h \rightarrow 1$, 
$P_{i,a}=P_{i,b}=P_{i,c}=P_{i,d}$, 
$P'_{i,a}=P'_{i,b}=P'_{i,c}=P'_{i,d}$ ($i=1,2$), $u=0$ and eq. 
(\ref{eq:finalmatrix}) becomes the standard undeformed 
result $^{\cite{tdlee}}$.

In order to compute the matrix element of $S$ to second
order in $\lambda$ we must generalize Wick's theorem since
the Feynman propagator in this case depends on the number
operator. We hope to report on this computation in a near 
future.

\section{Towards a multi-particle QFT}

In the last section we constructed the propagator and the
matrix element of the $S$-matrix to first order of the
coupling constant of a $q$-deformed spin-$0$ QFT. We stress
that we considered a $q$-deformed Heisenberg algebra and
the quantum mechanics was taken as being the standard one.
The advantage of studying a deformed system is that we
can make a contact with the non-deformed one in all
steps of the computation.

In this section we are going to discuss the possibility
of constructing a QFT interpreting each state
of a generalized Heisenberg algebra as a
particle with a different mass.  
We consider a Heisenberg algebra, generators of which are
given as
\begin{eqnarray}
A^{\dagger}  &=& S(P) \, \bar{T} \,\,  ,
\label{eq:real3} \\
A &=& T \, S(P) \,\,  ,
\label{eq:real4} \\
J_0 &=& \sqrt{V(P)+m^2_q} \,\, ,
\label{eq:real5} 
\end{eqnarray}
where $V(x)$ is a general function of $x$, and $T$,
$P$ are defined in eqs. (\ref{eq:abarra}, \ref{eq:momentum}), 
respectively, and $S(P)^2=J_0- m_q$. In the
special case where $V(P)=P^2$ we have the relativistic square-well 
algebra discussed in \cite{campos}.

Let us associate to each point of the discrete $\vec{k}$-space, 
defined in the previous section, an independent copy of 
the one-dimensional momentum lattice defined in section 3 
for each point of this $\vec{k}$-lattice  
so that $P_{\vec{k}}^{\dagger} = P_{\vec{k}}$ and 
$T_{\vec{k}}$, $\bar{T}_{\vec{k}}$ and $S_{\vec{k}}$ 
are defined by means of the previous definitions,
eqs. (\ref{eq:a}-\ref{eq:abarra} and \ref{eq:defS}),
through the substitution $P \rightarrow P_{\vec{k}}$. 
Then, we can define for each point of this lattice
a Heisenberg algebra as
\begin{eqnarray}
A_{\vec{k}}^{\dagger}  &=& S(P_{\vec{k}}) \, 
\bar{T_{\vec{k}}} \,\,  ,
\label{eq:real6} \\
A_{\vec{k}} &=& T_{\vec{k}} \, S(P_{\vec{k}}) \,\,  ,
\label{eq:real7} \\
J_0(\vec{k}) &=& \sqrt{{\vec{k}}^2+V(P_{\vec{k}})+m^2_q} \,\, ,
\label{eq:real8} 
\end{eqnarray}
where $S(P_{\vec{k}})^2=J_0(\vec{k})-\sqrt{{\vec{k}}^2
+m^2_q}$.

The Hilbert space of the associated QFT is spanned
by the vectors
\begin{equation} 
|0 \rangle, \,\, A^{\dagger}_{\vec{k}} |0 \rangle, \,\,
A^{\dagger}_{\vec{k}} A^{\dagger}_{\vec{k}'} |0 \rangle \,\,
\mbox{for} \,\, \vec{k}\not= \vec{k}', \,\,
(A^{\dagger}_{\vec{k}})^2 |0 \rangle, \,\, \cdots
\label{eq:hilbert} 
\end{equation}
Notice that the state $(A^{\dagger}_{\vec{k}})^n |0 \rangle$ 
has $J_0(\vec{k})$ eigenvalue given by \\
$\sqrt{{\vec{k}}^2+V(a n)+m^2_q}$, being interpreted as
the energy state of a particle with mass $\sqrt{V(a n)+m^2_q}$.
Thus, the associated interacting QFT would describe particles
with mass spectrum $\sqrt{V(a n)+m^2_q}$, with $n=1, 2, \cdots$,
unified by the generalized Heisenberg algebra under
consideration. The possibility of having a QFT unifying
a spectrum of particles of different masses is appealing 
with potential application in hadronic phenomenology. 
There are some points to be understood before developing 
such QFT as for instance Lorentz invariance, that plays an important 
role in a theory describing relativistic particles. We hope to 
develop this point and to report on such QFT in a near future. 

\vspace{0.4 cm}

\noindent
{\bf Acknowledgments:}
The authors thank CNPq/Pronex for partial support.

\newpage


\begin{thebibliography}{30}
\bibitem{algebra1} E. M. F. Curado and M. A. Rego-Monteiro, 
         Phys. Rev. {\bf E 61} (2000) 6255.
\bibitem{jpa} E. M. F. Curado and M. A. Rego-Monteiro, 
J. Phys. {\bf A 34} (2001) 3253. 
\bibitem{qosc} A. J. Macfarlane, J. Phys. {\bf A 22} (1989) 4581;
L. C. Biedenharn, J. Phys. {\bf A 22} (1989) L873.
\bibitem{comhugo} E. M. F. Curado, M. A. Rego-Monteiro and 
H. N. Nazareno, Phys. Rev. {\bf A 64} (2001) 12105; hep-th/0012244.
\bibitem{campos} M. A. Rego-Monteiro and E. M. F. Curado, ``Construction 
of a non-standard quantum field theory through a generalized Heisenberg 
algebra", preprint cbpf-nf-004/01; to appear in Int. J. Mod. Phys. {\bf A}.
\bibitem{circulo} M. A. Rego-Monteiro, Eur. Phys. J. {\bf C 21} (2001) 749.  
\bibitem{bonatsos} D. Bonatsos, J. Phys. {\bf A 25} (1992) L101;
D. Bonatsos, C. Daskaloyannis and A. Faessler, J. Phys. {\bf A 27} 
(1994) 1299.
\bibitem{dimakis1} A. Dimakis and F. Muller-Hoissen, Phys. Let.  
{\bf B 295} (1992) 242.
\bibitem{dimakis2} A. Dimakis, F. Muller-Hoissen and T. Striker, 
Phys. Let.  {\bf B 300} (1993) 141.
\bibitem{gensu2} E. M. F. Curado and M. A. Rego-Monteiro,
Physica {\bf A 295} (2001) 268; E. M. F. Curado and M. A. 
Rego-Monteiro, in preparation.
\bibitem{tdlee} See for instance: T. D. Lee, ``Particle Physics and
Introduction to Field Theory", Harwood academic publishers,
New York, 1981.

\end{thebibliography}
\end{document}